\documentclass[pra,showpacs,twocolumn,preprintnumbers,amsmath,amssymb,superscriptaddress]{revtex4}
\usepackage{graphicx}
\usepackage{dcolumn}
\usepackage{bm}

\begin{document}


\title{Quantum Hall effect with small numbers of vortices in Bose-Einstein condensates}

\author{Tim Byrnes}
\affiliation{New York University, 1555 Century Ave, Pudong, Shanghai 200122, China}
\affiliation{NYU-ECNU Institute of Physics at NYU Shanghai, 3663 Zhongshan Road North, Shanghai 200062, China}
\affiliation{National Institute of Informatics, 2-1-2 Hitotsubashi, Chiyoda-ku, Tokyo 101-8430, Japan}

\author{Jonathan P. Dowling}
\affiliation{Hearne Institute for Theoretical Physics, Department of Physics \& Astronomy, Louisiana State University, Baton Rouge, Louisiana 70803-4001, USA}

\date{\today}

\begin{abstract}
When vortices are displaced in Bose-Einstein condensates (BEC), the Magnus force gives the system a momentum transverse in the direction to the displacement. We show that Bose-Einstein condensates (BEC) in long channels with vortices exhibit a quantization of the current response with respect to the spatial vortex distribution.  The quantization originates from the well-known topological property of the phase around a vortex --- it is an integer multiple of $ 2 \pi $.  In a similar way to the integer quantum Hall effect, the current along the channel is related to this topological phase, and can be extracted from two experimentally measurable quantities: the total momentum of the BEC and the spatial distribution.  The quantization is in units of $ m/2h $, where $ m $ is the mass of the atoms and $ h $ is Planck's constant.  We derive an exact vortex momentum-displacement relation for BECs in long channels under general circumstances.  Our results presents the possibility that the configuration described here can be used as a novel way of measuring the mass of the atoms in the BEC using a topological invariant of the system. If an accurate determination of the plateaus are experimentally possible, this gives the possibility of a topological quantum mass standard and precise determination of the fine structure constant.    
\end{abstract}

\maketitle

\section{Introduction}
\label{sec:introduction}

One of the striking aspects of the integer QHE is the very precise quantization of the transverse conductance in units of $ e^2/h $, where $ e $ is the electronic change and $ h $ is Planck's constant.  This has relative uncertainties typically smaller than $10^{-10} $ between different samples and plateaus, which is an unprecedented level of accuracy for semiconductor systems which usually have unavoidable sources of disorder \cite{doucot04}.  The origin of the precision is now understood to be due to the Hall conductance being a topological quantity related to the Chern number \cite{hatsugai97,avron03}.  Understanding topological states of matter continues to gain importance, where there is currently an intense effort to investigating topological insulators \cite{hasan10} and applying these concepts to quantum computing using topological error correction codes, which have the highest error thresholds to date \cite{kitaev03,fowler09}.

\begin{figure}[ht!]
\begin{center}
\includegraphics[width=\columnwidth]{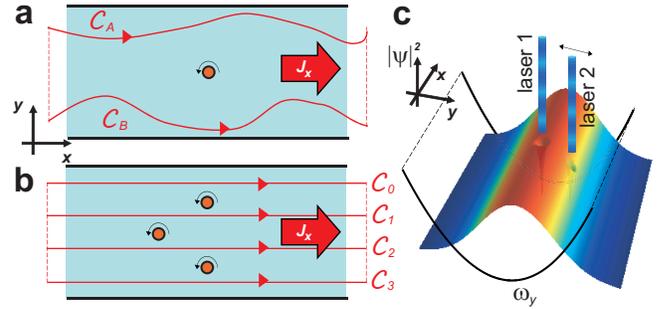}
\caption{\label{fig1}(Color online)  
Schematic configuration considered in this paper.  (a)(b) A Bose-Einstein condensate (BEC) is confined in a long channel along the $ x $-direction with one or several vortices (each marked with a circle, rotation orientation as marked) present in the central region.  Depending on the position of the vortices, the condensate flows with a total current $ J_x $. Contour integrals for various paths $ \cal C $ as discussed in the main text are marked. Dashed lines are contour integrals that are at $ x = \pm \infty $ which contribute zero.  (c) The experimental configuration considered in this paper.  We assume that the density profile along the channel is independent of $ x $, with the exception of density dips corresponding to vortex cores. Vortices are produced by optically based techniques such as stirring  or giving angular momentum to the BEC with Laguerre-Gauss modes.  Blue-detuned lasers pin the vortices and move them to various locations of the BEC.  }
\end{center}
\end{figure}

Realizing the QHE, and related topological states of matter, in systems other than semiconductors has therefore become an important pursuit in several fields. For Bose-Einstein condensates (BECs), there is a well-known equivalence between a magnetic field and rotation that allows for applying a vector potential to charge neutral atoms \cite{viefers08,fetter08}.  
%
%
%
%
This equivalence has suggested that the QHE may be accessed using BECs, when the bosons occupy the lowest Landau level.  Due to the interacting nature of the atoms in the BEC, it has also been predicted that the bosonic version of the fractional QHE should be observable, complete with non-Abelian quantum states.  Experimentally, among technical challenges such as heating, the difficulties are to precisely match the rotation frequency to the trapping frequency, which has so far prevented realizing the QHE in cold atoms.  In addition, there has been a large amount of interest in realizing synthetic magnetic fields, and more generally, gauge fields \cite{dalibard11}.  Experimentally this was realized in Bose-Einstein condensates with the production of vortices without rotation \cite{lin08}.  Several proposals for using such artificial magnetic fields to realize the fractional QHE \cite{sorensen05,hazefi07}, anomalous QHE \cite{liu10}, 
and quantum spin Hall effect \cite{goldman10}.  Some of the remarkable progress towards realizing such schemes experimentally include the demonstration of the superfluid Hall effect \cite{leblanc11}, spin Hall effect \cite{beeler13}, and measurement of the Chern number in Hofstadter bands \cite{aidelsburger14}.

In this paper we present an alternative and very simple approach to observe integer QHE behavior in a BEC.  
Our scheme does not possess an strict mathematical equivalence to the QHE as the approach as described in Refs. \cite{viefers08,fetter08} (or the bosonic version of it).  Nevertheless, it possesses several essential characteristics in common.  The quantization occurs with respect to the same observable as the standard QHE --- the current response of the condensate --- and occurs in units of $ m/2h $, where $ m $ is the atomic mass.  Furthermore, this quantization can be shown to originate from a topological quantity related to the phase of the wavefunction in the presence of vortices.  In the standard QHE it is known that the quantization occurs due to presence of vortices in the Brillioun zone \cite{hatsugai97}, in this respect the origin of our effect is the same, except that the vortices are in real space. If we consider the origin of the vortices as rotation, we obtaining a striking similarity to the standard conductance-magnetization quantization curve seen in the QHE, with discrete plateaus in the current response crossing over to linear behavior in the limit of many vortices.

\section{Vortex displacement-current relation}
\label{sec:theory}

\subsection{Single vortex configuration}
\label{sec:single}

Figure \ref{fig1} shows the basic configuration that we consider in this paper. The BEC is assumed to exist in a long channel, with the channel running in the $ x $-direction.  A strong $ z$-confinement allows us to consider an effectively two-dimensional system, such that the dynamics are entirely in the $x$-$y$ plane. The $ x $-confinement should be very weak, such that a net current can freely flow in the $ x $-direction. We furthermore assume that a small number of vortices are present in the center of the BEC.  Let us now consider the total current flowing in the $ x $-direction, equal to 
total velocity of all the atoms in the condensate in the $ x $-direction,
\begin{align}
J_x & = \frac{\langle p_x \rangle}{m} \nonumber = \int d \bm{r} j_x(\bm{r})
\end{align}
where $m$ is the mass of the atoms, $ \bm{j}(\bm{r})  = -\frac{i \hbar}{2m} \left[ \psi(\bm{r})^* \bm{\nabla} \psi(\bm{r}) -  \psi(\bm{r}) \bm{\nabla} \psi(\bm{r})^* \right] $, and $ \psi(\bm{r}) $ is the order parameter of the BEC. The primary assumption that we make is that due to the BEC being present in the long channel, it has no density dependence in the $ x $-direction, i.e. 
\begin{align}
\label{assumption}
\psi(x,y) = \sqrt{\rho (y)} e^{i S(x,y)},
\end{align}
where $ \rho (y) $ is a real function representing the density of the order parameter, and $ S(x,y) $ is the phase.  No assumptions are made about the phase distribution.  Strictly speaking, the presence of vortices already violates the assumption of $ x $-independence in the density as this implies local zeros in the condensate density. However, if the area occupied by the vortices is small relative to the total area, we show in Appendix \ref{sec:corrections} that this is a negligible contribution to the total current. 

Using the fact that the current is $ \bm{j} = \frac{\hbar}{m} \rho \bm{\nabla} S $, the total current $ J^x $ can then be evaluated to be
\begin{align}
\label{jxcurrentexpression}
J_x = \frac{\hbar}{m} \int_{-\infty}^\infty dy \rho(y) \int_{-\infty}^\infty dx \frac{\partial S(x,y)}{\partial x} .
\end{align}
The crucial observation is that the integral in the $ x $ direction may be evaluated exactly, as the density is independent of $ x $.  This follows from the well-known general topological property of the velocity $ \bm{v} (\bm{r}) =  \frac{\hbar}{m} \bm{\nabla} S (\bm{r}) $ in a BEC.  For an arbitrary path around $ N $ vortices, the line integral is \cite{pitaevskii03}. 
\begin{align}
\label{contour}
\oint \bm{v} \cdot d \bm{l} = \frac{\hbar}{m} \oint \bm{\nabla} S \cdot d \bm{l} = 2 \pi N \frac{\hbar}{m}  .
\end{align}
Now consider our particular geometry with a single vortex. For an arbitrary contour that extends to $ x = \pm \infty $ such as that shown in Fig. \ref{fig1}a, the contour integral can be written 
\begin{align}
\label{contoursplit}
I_{{\cal C}_B}  - I_{{\cal C}_A} = 2\pi
\end{align}
where 
\begin{align}
I_{\cal C} = \int_{\cal C} \bm{\nabla} S \cdot d \bm{l},
\end{align}
and $ {\cal C} $ is a contour that runs from $ x \in (-\infty,\infty) $ but may take any path along the way.   $ {\cal C}_A $ is a contour that runs above the vortex,  $ {\cal C}_B $ runs below the vortex.  The contributions from the edges at $ x= \pm \infty $ (dashed lines in Fig. \ref{fig1}a) give zero contribution as these are perpendicular to the center of mass momentum $ k_0 $ of the BEC in the $ x $-direction, and are very far from the vortex.  From the symmetry of the configuration we deduce that 
\begin{align}
I_{{\cal C}_A} & = k_0 - \pi \nonumber \\
I_{{\cal C}_B} & = k_0 +\pi,
\label{singlevortexphase}
\end{align}
where $ k_0 $ is the center of mass momentum of the condensate in the $ x $-direction. For the specific case of a contour that is a straight line parallel to the $ x $-axis, we have
\begin{align}
\int_{-\infty}^\infty dx \frac{\partial S(x,y)}{\partial x} = k_0  \pm \pi ,
\end{align}
where 
the sign is positive if $  y<y_1  $ and negative if $ y> y_1 $, and $ y_1 $ is the $y$-coordinate of the vortex.  

We may now evaluate (\ref{jxcurrentexpression}) directly to obtain
\begin{align}
J_x =  \frac{\hbar k_0 {\cal N}}{m}+ \frac{h}{2m} A_y .
\label{onevortexjx}
\end{align}
Here $ {\cal N} $ is the total number of particles in the BEC and the spatial asymmetry parameter is
\begin{align}
A_y = \int_{-\infty}^{y_1} dy \rho(y) - \int_{y_1}^\infty dy \rho(y). 
\label{ayonevortex}
\end{align}
The first term in Eq. (\ref{onevortexjx}) is a trivial overall offset to the current and is independent of the vortex component.  The second term however results entirely from the topological phase of the vortex. The physics described by (\ref{onevortexjx}) is rather simple: for various vortex displacements in the $ y$-direction in a condensate, a current in the $ x $-direction proportional to the parameter (\ref{ayonevortex}) is produced.  $ A_y $ is a parameter that counts the difference between the number of particles above and below the vortex.

\subsection{Multi-vortex configuration}
\label{sec:multi}

This argument is easily generalized to the case with multiple vortices.  Using the labeling for the contour integrations as shown in Fig. \ref{fig1}b, we obtain a system of equations satisfying 
\begin{align}
I_{{\cal C}_k} - I_{{\cal C}_l} = 2(k-l) \pi ,
\end{align}
where $ k,l \in [0,N] $, $ N $ is the number of vortices, and we have again assumed that all the vortices are located far away from the boundaries such that the contours at $ x = \pm \infty $ do not contribute. This can be solved to give   
\begin{align}
\label{generalphase}
I_{{\cal C}_l} = k_0 + (2l-N) \pi .
\end{align}
This allows us to write the current-asymmetry relation more generally for the multi-vortex case
\begin{align}
\label{currentasymmetry}
J_x = \frac{\hbar k_0 {\cal N}}{m} + G_{xy} A_y , 
\end{align}
where we have defined the Hall conductance-like quantity 
\begin{align}
\label{conductance}
G_{xy} \equiv \frac{h}{2m} N .
\end{align}
It is clear that this is quantized in units of  $ \frac{h}{2m} $ for each vortex that is present. Here the asymmetry parameter is 
\begin{align}
\label{genasymmetry}
A_y = \sum_{l=0}^N \frac{2l-N}{N} \int^{y_{l+1}}_{y_{l}} dy \rho(y)
\end{align}
where $ y_k $ is the $y$-coordinate of the $k$th vortex and we have defined $ y_0 \equiv \infty $ and $ y_{N+1}  \equiv -\infty $. 
The power of a relation such as (\ref{currentasymmetry}) --- as is also true for the QHE in semiconductors --- is that all the measurable quantities are easily accessible yet lead to a non-trivial quantum property of the system.

\subsection{Connections to Laughlin's gauge argument}
\label{sec:laughlin}

In the previous sections we have derived a connection between the net current flowing in a BEC with displacements of vortices.  This was derived from the topological integral of the phase around a vortex.  This can be viewed also from the point of view of the Magnus force when moving the vortex \cite{ao93,thouless96}. For the homogenous case
one may use an adaptation of Laughlin's gauge argument \cite{paramekanti04,laughlin81,huber11} to derive (\ref{onevortexjx}) in a limiting case.  


Let us first derive the current-asymmetry relation for infinitesimal displacements of the vortex.  Starting from (\ref{onevortexjx}), consider moving the vortex from $ y_1 $ to $ y_1 + \delta y $.  The change in the current is
\begin{align}
\delta J_x & = \frac{h}{2m} \left[A_y (y_1+ \delta y) - A_y (y_1) \right] \nonumber \\
& = \frac{h}{m} \int_{y_1}^{y_1  + \delta y} dy \rho(y) \nonumber \\
& \approx \frac{h}{m} \delta y \rho(y_1) .
\label{deltajx}
\end{align}
Associating the local density $ n = \rho(y_1) $ and the momentum $ J_x = p_x/m $, we obtain the relation
\begin{align}
\delta p_x \approx 2\pi \hbar n \delta y .
\label{onevortexapprox}
\end{align}
Using a similar argument one may derive the current for a multi-vortex configuration%
\begin{align}
\delta J_x & \approx \frac{h}{m} \delta y  \frac{1}{N} \sum_{l=1}^N \rho(y_l) .
\end{align}
For the homogenous case with $ \rho(y_l) = n $ this reduces to (\ref{onevortexapprox}).

On the other hand, we may obtain a similar relation using a modification of Laughlin's gauge argument (see for example 
Sec. IXB of Ref. \cite{paramekanti04} for a discussion of this).  Extend the two dimensional $ x$-$y$ plane to a torus and consider threading a flux in the $ y $ direction. The dimensions in the $ x$ and $ y $ directions are considered to be $ L_x $ and $ L_y $ respectively.  This may be achieved by creating a vortex-antivortex pair and moving them apart in the $ y $ direction until they annihilate at the opposite side of the torus.  The movement of the vortices in the $ y $ direction eventually give a momentum to the whole system in the $ x $ direction, which is the effect we are interested in.  The phase in the $ x $ direction that results from the flux threading is constrained to be $ e^{2\pi i x/L_x} $, i.e. the momentum given to the system is 
\begin{align}
\Delta p_x & = \frac{2 \pi \hbar {\cal N}}{L_x} \nonumber \\
& = 2 \pi \hbar n L_y 
\label{laughlinpx}
\end{align}
where $ n = \frac{{\cal N}}{L_x L_y } $. Integrating (\ref{onevortexapprox}) gives (\ref{laughlinpx}).  

While the Laughlin gauge argument gives qualitative agreement to our results, there are also several differences.  Firstly in Laughlin's argument one generally considers the momentum difference before and after the flux is threaded.  Our relations describe the current relation for an arbitrary vortex configuration. This is desirable particularly when a strict proportionality would like to be extracted, as is our case where $ m/h $ is the quantity that requires estimation.  Secondly, our results do not assume homogeneity in the $ y $-direction, which may be important to include in realistic BECs.

\begin{figure}
\begin{center}
\includegraphics[width=\columnwidth]{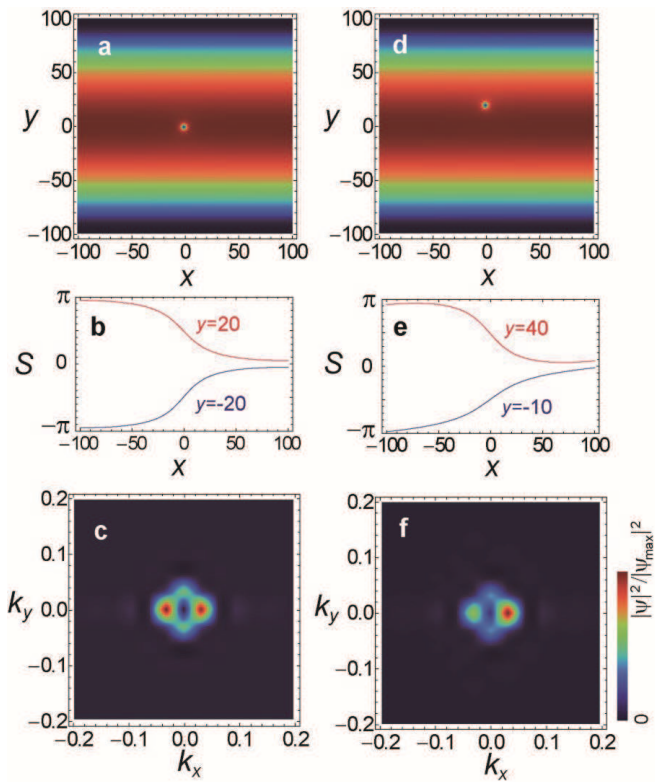}
\caption{\label{fig2} (Color online)  Stationary solutions of the Gross-Pitaevskii equation with a single vortex at $ \bm{r}_1 = (0,\delta y) $ with (a)(b)(c) $ \delta y = 0 $ 
(d)(e)(f)  $ \delta y = 20 $.  Plots for the condensate (a)(d) density, (b)(e) phase, (c)(f) momentum distribution are shown. Length scales are in units of the healing length $ \xi = \sqrt{\frac{\hbar^2}{2m g n}} $ where $ n $ is the maximum density of the BEC, along $ y = 0 $. We assume parameters $ \hbar \omega_y/E_0 =0.02 $, $ V_0/E_0 =1 $, $k_0=0 $, where the energy scale is $ E_0 = \frac{\hbar^2}{2m \xi^2} = gn $.  The delta potential is located at the vortex position $ \bm{r}_1 $.  }
\end{center}
\end{figure}

\section{Numerical simulations}
\label{sec:numerical}

To illustrate the effect, we perform numerical simulations of the Gross-Pitaevskii equation
\begin{align}
i \hbar \frac{\partial \psi}{\partial t} = \left[ - \frac{\hbar^2 \nabla^2 }{2m} + \frac{1}{2} m \omega_y^2 y^2 + V_0 \sum_k \delta(\bm{r}-\bm{r}_k) + g |\psi |^2 \right] \psi, \label{gpequation}
\end{align}
where $ \omega_y $ is the trapping frequency in the $ y $-direction, $ V_0 $ is the strength of the pinning potentials at locations $ \bm{r}_k $, and $ g $ is the interaction strength. Using a real space second order finite difference time decimation (FDTD) method on a $100 \times 100 $ site grid simulating the region $ x\in [-x_{\mbox{\tiny max}},x_{\mbox{\tiny max}}], y\in [-y_{\mbox{\tiny max}},y_{\mbox{\tiny max}}] $ evolving in time using the backwards Euler method.  In order to obtain stationary vortex states we evolve in imaginary time starting from an approximate vortex wavefunction.  In order to have stable vortex solutions under imaginary time evolution, we pin the vortex using a local delta function at the desired vortex position, which has a negligible effect on the phase and density of the condensate wavefunction. To support a constant current in the $ x $-direction, we employ (anti-)periodic M{\"o}bius boundary conditions $ \psi(-x_{\mbox{\tiny max}},y) = (\pm 1)^N \psi(x_{\mbox{\tiny max}},-y) $ for an even (odd) number of vortices.  The phase factor of $ (\pm 1)^N $ is necessary as each vortex flips the phase by either $ \pi $ going from $ x = -\infty \rightarrow \infty $, as discussed in the main text.  The M{\"o}bius boundary conditions are necessary as the current far to the left of the vortices are predominantly in the $ +y$ direction, whereas far to the right the current is in the $ -y $ direction, or vice versa. Parameters are chosen consistent with BECs with large particle numbers where the condensate radius is much larger than the healing length \cite{streed06}.

Our results for a single vortex are shown in Fig. \ref{fig2}. The real space, phase, and momentum space distributions for two vortex positions are shown.  In both cases the imaginary time evolution ensures a
quiescent stationary state of the BEC which smooths out the density fluctuations in the $ x $-direction. The lack of such density variations is our primary assumption, and is the corresponding situation in 
the standard QHE to an equilibrium state free of transient dynamics.  The real space images [Fig. \ref{fig2}(a)(d)] show that such a state is achieved with a stable vortex pinned at their respective positions. The phase variations [Fig. \ref{fig2}(b)(e)] agree well with the discussion relating to Fig. \ref{fig1}(a), showing that there is a phase change of $\pm \pi $ depending on whether the $ y $-position is above or below the vortex. For the central vortex position the momentum distribution is symmetrically distributed such that the average momentum (and hence current) is $ \langle k_x \rangle \approx 0 $.  For the displaced vortex position, the momentum distribution shifts to the right, indicating a non-zero $ \langle k_x \rangle > 0 $, as predicted by the relation (\ref{onevortexjx}). We emphasize here that the center-of-mass momentum is $ k_0 =0 $, so that the net momentum in the $ x $-direction in (\ref{onevortexjx}) results entirely from the vortex displacement. 

\begin{figure}[t]
\begin{center}
\includegraphics[width=\columnwidth]{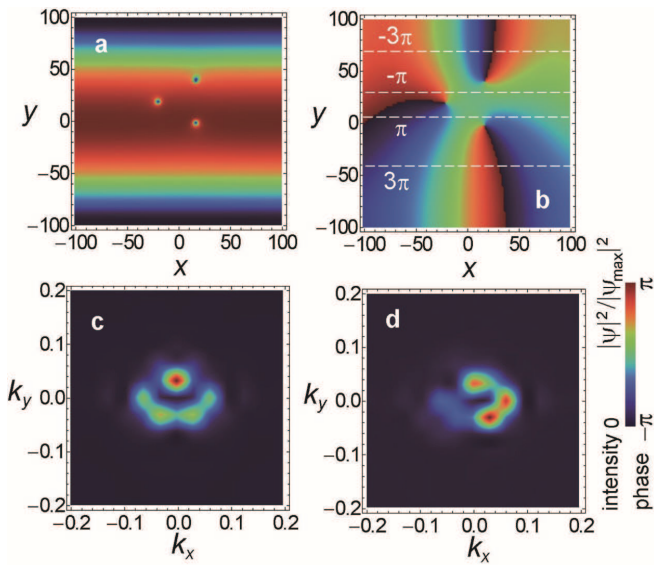}
\caption{\label{fig3}(Color online)  Stationary solutions of the Gross-Pitaevskii equation with three vortices with coordinates $\bm{r}_1 = (10\sqrt{3},20+\delta y) $, $\bm{r}_2 = (-10\sqrt{3},\delta y) $, $ \bm{r}_3 = (10\sqrt{3},-20+ \delta y)$,  where  (a)(b)(d) $ \delta y=20 $ and (c) $ \delta y=0 $.  Plots for the condensate (a) density, (b) phase, and (c)(d) momentum distribution are shown. The same parameters and units as Fig. \ref{fig2} are used. }
\end{center}
\end{figure}

\begin{figure}[t]
\begin{center}
\includegraphics[width=\columnwidth]{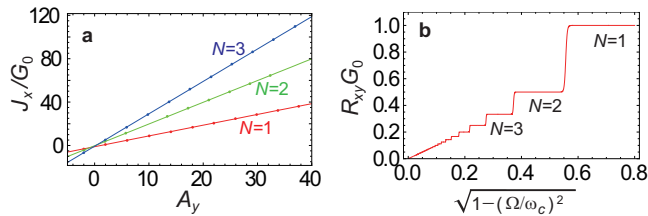}
\caption{\label{fig4}(Color online) (a) Total current $ J_x $ versus asymmetry parameter $ A_y $ for various vortex number $ N$. Points show the numerically evaluated values and lines are fits to the data.  The current $ J_x $ is calculated in units of $ G_0 = \frac{h}{2m} $, the elementary unit of conductance, such that the gradient should be an integer.  A linear fit to the data gives a gradient of $ 0.99$, $ 2.00 $,  $2.99 $ for the $ N = 1,2,3 $ vortex cases respectively. (b) Resistance $ R_{xy} = 1/G_{xy} $ versus the rotation parameter $ \sqrt{1- (\Omega/\omega_c)^2} $. Resistance is measured in units of inverse $ G_0 $ and $ \kappa = 1 $ is used. The number of vortices generated by the rotation are labeled. Plateaus are rounded to account for the uncertainty in vortex number between plateaus.  }
\end{center}
\end{figure}

Similar results are obtained for the multi-vortex case as shown in Fig. \ref{fig3}.  The same procedure as the single vortex case is repeated for a configuration of three vortices in an equilateral triangle, displaced by various $ y $ positions. The phase relation (\ref{generalphase}) can be seen to hold by taking lines at various $ y $-positions, giving phase shifts of $-3\pi, -\pi,\pi, 3 \pi $.   The momentum distributions again shift towards the positive $ \langle k_x \rangle >0 $ direction once the vortices are displaced.  The current-asymmetry relation for various numbers of vortices and $ y$-displacements are shown in Fig. \ref{fig4}a.  We see a perfect proportionality relation as predicted by (\ref{currentasymmetry}). A linear fit to the data points give quantization to the integer multiples of $ G_0 \equiv h/2m $.  While good agreement with the theory is observed, we attribute discrepancies to exact integral quantization to the relatively short channel that we use in the simulations, of which the length is only of the order of the width.  The very precise quantization as seen in the QHE is a result of the whole system contributing to the conductance. Thus as the system size is increased we expect the precision to improve (see also Appendix \ref{sec:corrections}).  In experimental systems we expect that much longer channels can be produced, and so can benefit from this scaling effect. As our arguments are based on rather general topological considerations of the phase around vortices in a BEC, the effects of local disorder are not detrimental to the effect, assuming that they do not cause large variations in the density along the channel.

We now show the explicit quantization relation of the inverse conductance $ R_{xy} = 1/ G_{xy} $ to show most clearly the analogy with the QHE. Let us assume that the vortices are originally produced by a rotation of frequency $ \Omega $.  The number of vortices that are produced in two dimensions can then be estimated to be  \cite{kato11}
\begin{align}
N =  \left\lfloor \kappa \frac{\Omega/\omega_c}{\sqrt{ 1 - (\Omega/\omega_c)^2 }} \right\rceil
\label{vortexnumber}
\end{align}
where $ \omega_c $ is the critical frequency which gives rise to the proliferation of vortices, $ \kappa  $ is a dimensionless proportionality constant, and the bracket $ \lfloor  \rceil $ rounds to the nearest integer. 
Let us now consider that in this situation we examine the $ J_x $ current and vortex positions $ y_l $ in the rotating frame.  Using this we may then calculate a conductance quantity $ G_{xy} $ as the vortex positions are changed.  
Taking this value as the vortex number in $ G_{xy} $, we obtain a curve which is remarkably reminiscent of the conductance-magnetic field relation in the QHE (Fig. \ref{fig4}(b)).  For slow rotations we recover the conductance plateaus corresponding to low vortex numbers.  For fast rotations there is a proliferation of vortices, and simultaneously the resistance $ R_{xy} $ diminishes as $ \propto 1/N $, which gives a linear relation.  We note that a  similar curve can also be obtained by plotting $ 1/\Omega $ versus $ R_{xy} $, although this gives a square root relation as the critical $\Omega/\omega_c \rightarrow 1 $  is approached. In practice, it is likely that directly rotating the BEC is not the best experimental method for vortex generation (see Sec. \ref{sec:exp}).  Nevertheless, as rotations are the corresponding quantity to the magnetic field in the BEC case, the equivalence to the standard QHE is most clearly illustrated by the conductance quantization as shown in Fig. \ref{fig4}(b).

\section{Experimental realization}
\label{sec:exp}

We finally discuss the likely experimental configuration of our proposal. A oblate atomic BEC with trapping frequencies satisfying $ \omega_z > \omega_y \gg \omega_x $ would be prepared, such as the dynamics would primarily occur in the $x$-$y$ directions. 
To ensure that such vortices can exist in the channel, the $ y $-confinement and density should be such that the width of the BEC is larger than the healing length. The schematic configuration is shown in Fig. \ref{fig1}(c). Starting from such a configuration, small numbers of vortices would be generated. This is most suitably done with optically based techniques such as stirring with a blue-detuned laser \cite{madison00,raman01,aboshaeer01}, or adiabatically introducing angular momentum to the BEC using Laguerre-Gauss modes \cite{andersen06,ryu07,nandi04}. In order to extract $G_{xy} $ from the current and asymmetry relation, a variety of vortex positions are required. To achieve this in the most controlled fashion, pinning of the vortices at a given location is desirable.  Vortex-antivortex pairs may be reliably generated and pinned at a desired location by the use of two blue-detuned lasers moved through the BEC \cite{samson12,samson15}. Once the vortices are generated, the density distribution of the BEC is estimated using high-resolution spatial imaging techniques such as scanning electron microscopy \cite{gericke07}.  Other {\it in-situ} methods such as as phase contrast imaging could be used to obtain a density distribution of the BEC to identify the vortex positions \cite{andrews96,ilookeke14}.  The total current $ J_x = \frac{\hbar }{m}\sum_{k_x,k_y}  k_x |\psi_{k_x k_y} |^2  $ can be extracted from the high-resolution velocity distribution of the BEC, which can performed by Bragg spectroscopy \cite{stenger99} or time-of-flight imaging \cite{pethick08}. 

In order to measure the conductance plateaus, one may directly use the relation (\ref{currentasymmetry}), or alternatively the differential form (\ref{deltajx}) which can be written as
\begin{align}
\delta \tilde{J_x} & = \frac{h}{m} \int_{y_1}^{y_1+ \delta y} dy \tilde{\rho} (y) \nonumber \\
& \approx  \frac{h}{m} \delta y \tilde{\rho} (y_1)
\label{normrelation}
\end{align}
where the current per particle is 
\begin{align}
\delta \tilde{J_x} \equiv \frac{J_x (y_1 + \delta y ) - J_x (y_1)}{\cal N}
\end{align}
and the normalized density is
\begin{align}
\tilde{\rho} (y) = \frac{\rho (y)}{\cal N} .
\end{align}
The advantage of writing the relations using normalized quantities is that they become insensitive to number fluctuations in the BEC which may originate from variations in the initial conditions or particle loss. Thus during the measurement process, only the relative position and momentum distributions are required. The ratio $ h/m  $ is then extracted by looking at the differential variation of the current $ \delta \tilde{J_x}  $ with variations of the vortex position $ \delta y $.  The spatial measurement then plays the role of obtaining $  \tilde{\rho} (y_1) $, i.e. the relative density of the condensate at $ y $-coordinate of the vortex.  This can also be obtained by interpolating the density without the vortex core.

\section{Applications}

Since the conductance $G_{xy}$ is quantized in units of $G_0$, our result presents the possibility of an novel method of measuring the mass of atoms in the condensate.  We note that this would be a mass spectrometer that would be able to measure the {\it absolute} mass of the atoms, rather than the {\it relative} atomic mass.  Currently Penning traps are the most precise mass spectrometers \cite{becker06} achieving relative uncertainties of typically $\sim 10^{-10}$ for the relative atom mass.  The (absolute) atomic mass unit is known to an larger uncertainty of $ \sim 10^{-8} $ \cite{mohr12}.  If an extremely precise measurement of absolute mass of an atom became possible, this would allow for a redefinition of the kilogram as a fixed number of atoms of a particular type; $^{87}\mbox{Rb}$ for instance (see Appendix \ref{sec:mass}). Since the conductance $ G_0 $ is the quantity that would be measured in our proposal, the mass would be measured in units of the Planck's constant. This is consistent with other methods that aim to contribute to a redefinition of the kilogram such as the watt balance and silicon atom counting methods.  As another potential application, the quantity $ h/m $ itself is of interest in the context of determination of the fine structure constant, by combining estimates of the Rydberg constant, the relative mass of an atom, and  $ h/m $ \cite{mohr12}.  The first two factors are determined to better than $ 10^{-10} $, while currently $ h/m $ can be estimated to a level of $ 10^{-8} $. 

The spatial and velocity distribution measurement methods both have a finite resolution which may appear to severely limit the precision of the quantities to be estimated in (\ref{normrelation}). However, bulk quantities such as $ \delta \tilde{J_x} $ scale well with finite resolution, e.g. for Simpson's rule the errors scale as the fourth power of the discretization. 
For $ \tilde{\rho} (y_1) $, assuming that the vortex is pinned by the blue-detuned laser so that it is always present at the same location, then one can measure the relative density with high accuracy by many repetitions of the experiment. 
To minimize the uncertainty of the density measurement it is advantageous to move the vortex in the vicinity of the maximum density of the BEC, where the derivative with position is zero, i.e. $ \tilde{\rho} (y_1 \pm \delta y) \approx \tilde{\rho} (y_1) \pm \frac{d \tilde{\rho}}{dy} \delta y $.  In this way errors due to finite resolution can be mitigated. This is also advantageous in terms of corrections to the current due to the vortex (see Appendix \ref{sec:corrections}).  Finally, $ \delta y $ can be set by the pinning laser which is controlled and hence is not measured directly in the BEC. 
We therefore estimate that the main sources of error will arise from density and thermal fluctuations in the BEC,  and other experimental issues such as calibration of measured and controlled quantities, and repeatability.  Although it is ultimately an experimental question of whether sufficiently low uncertainties can be attained, at the level of the system
the topological nature of the observable should make the measurement of $G_{xy}$ rather robust under a variety of conditions, in analogy with the 
standard QHE.

\section{Summary and conclusions}
\label{sec:conclusions}

In summary, we have shown an alternative method of investigating integer quantum Hall physics using BECs with small numbers of vortices.  While this approach does not have a precise mathematical equivalence of previous approaches \cite{viefers08,fetter08}, much of the essential physics is in common, where the topological phase around the vortices give rise to a quantized Hall conductance-like quantity. In both cases the current response along the channel is measured, but here the proportionality is with respect to an asymmetry parameter, as opposed to the standard QHE where it is the potential difference in the transverse direction.  Alternatively the density at the same $ y $-coordinate as the vortex can be measured instead of the asymmetry parameter.  A potential application is to use the quantization of the conductance in units of $ h/2m $ as a novel way of measuring the absolute mass of the atoms, using a topological invariant of the system. While in this paper we have implicitly assumed an atomic BEC, other types of BEC, such as exciton-polaritons \cite{deng10,byrnes14}, should also be suitable to observe the effect as the effect relies only on the topological phase of the condensate.

\begin{acknowledgments}
T.B. and J.P.D. thank Ricardo Carretero and Brian P. Anderson for discussions. T.B. would like to acknowledge support from the Inamori foundation, NTT Basic Research Laboratories, Shanghai Research Challenge Fund. J.P.D. would like to acknowledge support from the Air Force Office of Scientific Research, the Army Research Office and the National Science Foundation. 
\end{acknowledgments}

\appendix

\section{Corrections to the conductance quantization due to vortices}
\label{sec:corrections}

The main assumption made in our calculations is Eq. (3), that the BEC's density is uniform in $ x $-direction. The presence of vortices clearly violates this assumption. In this section we show to what extent the presence of vortices make to the final result. 

The condensate wavefunction with a single vortex can be written
\begin{align}
\label{assumptiongeneral}
\psi(x,y) = \left[ f(y)- \Delta f(r_{\mbox{\tiny v}} ) \right] e^{i S(x,y)},
\end{align}
where $ \Delta f(r_{\mbox{\tiny v}}) $ is the density deviation due to the presence of the vortex, $ f(y) = \sqrt{\rho(y)}$,  $ r_{\mbox{\tiny v}} = | \bm{r} -\bm{r}_1 |$, and the vortex is located at the position $ \bm{r}_1 $. This is typically a positive quantity that has a maximum at the vortex core and approaches zero at a distance of the order of the healing length.  The current can be written 
\begin{align}
\label{jxcurrentexpressionsuppl}
J_x = & \frac{\hbar}{m} \int dx dy \frac{\partial S}{\partial x} \left[f^2 (y) - 2 f(y) \Delta f(r_{\mbox{\tiny v}}) + (\Delta f(r_{\mbox{\tiny v}}))^2 \right]   .
\end{align}
Evaluating this expression as in the main text, we obtain
\begin{align}
J_x = \frac{\hbar}{m} \left[ k_0 {\cal N} + \frac{A_y}{2}  + \Delta J_x  \right],
\label{onevortexjxsuppl}
\end{align}
where we have used the same definitions as the main text and 
\begin{align}
\Delta J_x = \int dx dy \frac{\partial S}{\partial x} \Delta f(r_{\mbox{\tiny v}}) \left[\Delta f(r_{\mbox{\tiny v}}) 
-2 f(y) \right] . 
\end{align}
The main contribution to the integral is the region around the vortex, which is assumed to be a small region in comparison to the scale of the function $ f(y) $. Shifting the coordinates to the center of the vortex, we can then approximate the phase in this region as being $ e^{i (\phi + S_0 )}$ (here $S_0 $ is a constant), and $ f(y) \approx f_0 + f_1 y + \dots$ to first order. This gives $ \frac{\partial S}{\partial x} = - \frac{\sin \phi}{r_{\mbox{\tiny v}}} $, and hence
\begin{align}
\Delta J_x = & - \int dr_{\mbox{\tiny v}} (\Delta f(r_{\mbox{\tiny v}}))^2 \int d \phi \sin \phi \nonumber \\
&  + 2 f_0 \int dr_{\mbox{\tiny v}} \Delta f(r_{\mbox{\tiny v}}) \int d \phi \sin \phi \nonumber \\
&  + 2 f_1 \int dr_{\mbox{\tiny v}} r_{\mbox{\tiny v}} \Delta f(r_{\mbox{\tiny v}}) \int  d \phi  \sin^2 \phi + \dots
\end{align}
The first and second terms are zero due to the integral over the phase.  Thus for local densities $ f(y)$ that are flat give zero correction to the current.  The first order correction enters when there is a gradient $ f_1  $ in the local density.  The last term can be evaluated and we obtain 
\begin{align}
\Delta J_x = 2 \pi  f_1 \int dr_{\mbox{\tiny v}} r_{\mbox{\tiny v}} \Delta f(r_{\mbox{\tiny v}}) + \dots
\end{align}
Let us now estimate the order of magnitude of each of the terms in  (\ref{onevortexjxsuppl}). Let us write the average density of the BEC as $ n \sim \frac{{\cal N} }{L_x L_y } $, where $ L_x, L_y $ are the lengths of the BEC in the $ x, y $ directions. Taking the magnitudes of $ f_0 \sim \sqrt{n} $, $ f_1 \sim \sqrt{n}/l $, and $ \Delta f(r_{\mbox{\tiny v}}) \sim \sqrt{n} $, where $ l $ is the length scale associated with the gradient, we have
\begin{align}
J_x \sim \frac{\hbar n}{m} \left[ k_0 L_x L_y + L_y + \frac{{\cal A}_{\mbox{\tiny vortex}}}{l} + \dots \right]  .
\end{align}
where $ {\cal A}_{\mbox{\tiny vortex}} $ is the area of the vortex. For the correction due to current due to the vortex to be negligible, we thus require that 
\begin{align}
\label{conditions}
L_y \gg \frac{{\cal A}_{\mbox{\tiny vortex}}}{l}.  
\end{align}
This is satisfied if the area of the vortex is very small compared to the condensate, and if the vortex is present in a very flat region of the BEC.  

\section{Mass standards}
\label{sec:mass}

The aim of a mass standard is to create a new definition of the kilogram, which is currently defined as the mass of a platinum-iridium alloy artifact stored in France \cite{becker06}.  As with other definitions such as the meter and the second, it is desirable to use fundamental constants of nature rather than material artifacts, for several reasons such as stability, reproducibility across the world, and other practical issues causing the artifact's mass to drift in time.  We give some more detail on how our proposal would be connected with the kilogram mass standard. 

One approach to a mass standard is to define 1kg to be equal to a certain number of carbon-12 atoms.  However, there is already the definition that 1 mole ($=N_A$, Avogadro's number) of carbon-12 is 12 grams exactly. This means that when the kilogram is redefined
the constraint 
\begin{align}
\label{carbonconstraint}
N_A  m_C = 0.012
\end{align}
must also be satisfied, where $m_C$ is the mass of 1 carbon-12 atom (in kg).  Currently, both $N_A$ and $m_C$ are experimentally determined quantities.  In a redefinition of the kilogram using carbon atoms, the Avogadro constant would be fixed to a particular number, 
for example $ N_A = 6.02214 \times 10^{23} $ exactly. Then according to the constraint (\ref{carbonconstraint}) this fixes $m_C$ also exactly.  Alternatively, $m_C$ could also be fixed, and this would fix $N_A$ according to (\ref{carbonconstraint}). Fixing $m_C$ or $N_A$ is equivalent in this sense. 

A very precise measurement of the mass of the atoms can therefore equivalently contribute towards the mass standard.
While a BEC of carbon-12 is not practical, the relation (\ref{carbonconstraint}) can be converted to something more convenient for this purpose.  Introducing the mass of the atom species that undergoes BEC $ m $ (such as $^{87}$Rb) we have
\begin{align}
N_A \frac{m_C}{m} m	 = 0.012 .
\end{align}
The relative mass ratio $ \frac{m_C}{m} $ can be measured very precisely (typically $ 10^{-10} $ relative error) using Penning traps \cite{becker06}. In this way by measuring the (absolute) mass of the BEC atom, this can be used to determine the Avogradro constant, which defines the mass standard.


\end{document}